\documentclass[11pt]{biometrika}
\usepackage{amsmath,amsfonts,amssymb,latexsym}
\usepackage{graphicx}
\usepackage{times}
\usepackage{bm}
\usepackage{natbib}

\usepackage[plain,noend]{algorithm2e}

\makeatletter
\renewcommand{\algocf@captiontext}[2]{#1\algocf@typo. \AlCapFnt{}#2} 
\def\@algocf@capt@plain{top}
\renewcommand{\algocf@makecaption}[2]{%
  \addtolength{\hsize}{\algomargin}%
  \sbox\@tempboxa{\algocf@captiontext{#1}{#2}}%
  \ifdim\wd\@tempboxa >\hsize
    \hskip .5\algomargin%
    \parbox[t]{\hsize}{\algocf@captiontext{#1}{#2}}
  \else%
    \global\@minipagefalse%
    \hbox to\hsize{\box\@tempboxa}
  \fi%
  \addtolength{\hsize}{-\algomargin}%
}
\makeatother

\newcommand{\R}{\mathbb{R}}
\newcommand{\Cov}{\mathcal{C}}
\newcommand{\Exp}{E}
\newcommand{\psih}{\psi_{h}}
\newcommand{\Hs}{\mathcal{H}}
\newcommand{\psiT}{\psi_{h}^{I}}
\newcommand{\PsiT}{\Psi_{h}^{I}}

\newcommand{\XiL}{\tilde{\Xi}_{h}}
\newcommand{\PsiInv}[1]{\Psi_{h}^{-#1}}

\begin{document}




\title{Bayesian inference for partially observed SDEs driven by fractional Brownian motion}

\author{A. Beskos}
\affil{Department of Statistical Science, University College London,\\
1-19 Torrington Place, London WC1E 7HB, U.K. \email{a.beskos@ucl.ac.uk}}

\author{J. Dureau}
\affil{Department of Statistics, London School of Economics,\\
Houghton Street, London WC2A 2AE, U.K.
\email{dureau.joseph@gmail.com} }

\author{\and K. Kalogeropoulos}
\affil{Department of Statistics, London School of Economics,\\
Houghton Street, London WC2A 2AE, U.K.
\email{k.kalogeropoulos@lse.ac.uk} }

\maketitle

\begin{abstract}
We consider continuous-time diffusion models driven by fractional Brownian motion.
Observations are assumed to possess a non-trivial likelihood given the latent path.
Due to the non-Markovianity and high dimensionality of the latent paths, estimating posterior expectations is computationally challenging. We present a reparameterization framework based on the Davies and Harte method for sampling stationary Gaussian processes and use it to construct a Markov chain Monte Carlo algorithm that allows computationally efficient Bayesian inference. The algorithm is based on a version of hybrid Monte Carlo
that delivers increased efficiency when applied on the high-dimensional latent variables arising in this context. We specify the methodology on a stochastic volatility model, allowing for memory in the volatility increments through a fractional specification. The  methodology is illustrated on simulated data and on the S\&P500/VIX time series the posterior distribution favours values of the Hurst parameter, smaller than $1/2$, pointing towards medium range dependence.
\end{abstract}


\section{Introduction}

A natural continuous-time modeling framework for processes with memory uses fractional Brownian motion as the driving noise. This is a zero mean self-similar Gaussian process, say $B^{H} = \{B^H_t, t\ge 0\}$, of covariance $\Exp(B^{H}_s B^{H}_t) = (\, |t|^{2H}+|s|^{2H}-|t-s|^{2H})/2$, $0\le s\le t$,
parameterized by the Hurst index $H\in (0,1)$.
For $H=1/2$ we get the Brownian motion with independent increments. The case of $H>1/2$ gives smoother paths of infinite variation with positively autocorrelated increments that exhibit long-range dependence, in the sense that the  autocorrelations are not summable. For $H<1/2$ we obtain rougher paths with negatively autocorrelated increments exhibiting medium-range dependence; the autocorrelations are summable but decay more slowly than the exponential rate characterizing short-range dependence.

Since the  pioneering  work of \cite{mand:68}, various applications
have used fractional noise in models to capture self-similarity, non-Markovianity or sub-diffusivity and super-diffusivity; see for example \cite{kou:08}. Closer
to our context, numerous studies have explored the well-posedness of stochastic differential equations driven by $B^{H}$,
\begin{equation}
\label{eq:big}
dX_t = b(X_t)\,dt +\sigma(X_t)\,dB_t^H\ ,
\end{equation}
for given functions $b$ and $\sigma$; see \cite{biag:08} and references therein. Unlike most inference methods for models based on \eqref{eq:big} in non-linear settings, which have considered direct and high frequency observations on $X_t$ \citep{rao:10},
the focus of this paper is on the partial observation setting. We provide a general framework, suitable for incorporating information from additional data sources, potentially from different time scales. The aim is to perform full Bayesian inference for all parameters, including $H$, thus avoiding non-likelihood-based methods typically used in this context such as least squares. The Markov chain Monte Carlo algorithm we develop is relevant
in contexts where observations $Y$ have a non-trivial likelihood, say $p(Y\mid B^{H})$, conditionally on the driving noise.
We assume that  $p(Y\mid B^{H})$ is known and genuinely a function of the infinite-dimensional
latent path $B^{H}$, that is we cannot marginalize the model onto finite dimensions.
While the focus is on a scalar context, the method can in principle be applied to higher dimensions at increasing computational costs; e.g. with likelihood $p(Y\mid B_i^{H_i},i=1,\ldots, \kappa)$ for Hurst parameters $H_i$, $i=1,\ldots, \kappa$.

A first challenge in this set-up is the intractability of the likelihood function
\begin{equation*}
p(Y\mid\theta) =   \int p(Y\mid X,\theta)\, p(dX\mid \theta)\ ,
\end{equation*}
with $\theta\in \R^{q}$ denoting all the unknown parameters. A data augmentation approach is adopted, to obtain samples from the joint posterior density
\begin{equation*}
\Pi(X,\theta\mid Y) \propto p(Y\mid X,\theta)\,p(X\mid\theta)\,p(\theta) \ .
\end{equation*}
In practice, a time-discretized version of the infinite-dimensional path $X$ must be considered, on a time grid of size $N$. It is essential to construct an algorithm with stable performance as $N$ gets large, giving accurate approximation of the theoretical posterior $p(\theta\mid Y)$.

For the standard case $H=1/2$, efficient data augmentation algorithms, with mixing time not deteriorating with increasing $N$, are available \citep{rob:str01,goli:08,kal:rob:del10}. However, important challenges arise if $H\neq 1/2$. First, some parameters, including $H$, can be fully identified by a continuous path of $X$ \citep{rao:10}, as the joint law of $\{X,H\}$ is degenerate, with $p(H\mid X)$ being a Dirac measure. To avoid slow mixing, the algorithm must decouple this dependence. This can in general be achieved by suitable reparameterization, see the above references for $H=1/2$, or by a particle algorithm \citep{Andrieu2010}. In the present setting the latter direction would require a sequential-in-time realization of $B^{H}$ paths of cost $\mathcal{O}(N^2)$ via the \cite{hosk:81} algorithm or approximate algorithms of lower cost \citep{norr:99}.
Such a method would then face further computational challenges, such as overcoming path degeneracy and producing unbiased likelihood estimates of small variance. The method developed in this paper is tailored to the particular structure of the  models of interest, that of a change of measure from a Gaussian law in high dimensions. Second, typical algorithms for $H=1/2$ make use of the Markovianity of $X$. They exploit the fact that given $Y$, the $X$-path can be split into small blocks of time with updates on each block involving computations only over its associated time period. For $H\neq 1/2$, $X$ is not Markovian, so a similar block update requires calculations over its complete path. Hence, a potentially efficient algorithm should aim at updating large blocks.

In this paper, these issues are addressed in order to develop an effective Markov chain Monte Carlo algorithm. The first issue is tackled via a reparameterization provided by the Davies and Harte construction of $B^{H}$. For the second issue we resort to a version of the hybrid Monte Carlo  algorithm \citep{duna:87}, adopting ideas from \cite{besk:11,besk:14}. This algorithm has mesh-free mixing time, thus is particularly appropriate for big $N$.

The method is applied on a class of stochastic volatility models of importance in finance and econometrics. Use of memory in the volatility is motivated by empirical evidence \citep{DingEtAl1993, LobSav1998}. The autocorrelation function of squared returns is often observed to be slowly decaying towards zero, not in an exponential manner that would suggest short range dependence, nor implying a unit root that would point to integrated processes. In discrete time, such effects can be captured for example with long memory stochastic volatility model of \cite{BreEtAl1998}, where the log-volatility is a fractional autoregressive integrated moving average process. In continuous time, \cite{ComRen1998} introduced the model
\begin{align}
    dS_t&=\mu S_t dt + \sigma_S(X_t)\,S_t\,dW_t\,,     \label{eq:S}  \\
    dX_t&=b_X(X_t,\zeta)dt+\sigma_X(X_t,\zeta)\,dB_t^H\ , \,\,\,S_0>0\ , \, X_0=x_0\in \mathbb{R}\ , \quad  0\le t\le \ell \  .\label{eq:X}
\end{align}
Here, $S_t$, $X_t$ are the asset price and volatility processes respectively and $W$ is standard Brownian motion independent of $B^H$. The definition involves also the length $\ell>0$ of the time-period under consideration, and functions $\sigma_S:\R\mapsto \R$, $b_X:\R\times \R^{p}\mapsto \R$ and $\sigma_X:\R\times \R^{p}\mapsto \R$, together with unknown parameters $\mu\in \R$, $\zeta\in \R^p$, $p\ge 1$.
In \cite{ComRen1998} the log-volatility is a fractional Ornstein--Uhlenbeck process, with $H>1/2$, and the paper argues that incorporating long memory in this way captures the empirically-observed strong smile effect for long maturity times. In contrast with previous literature, we consider the extended model that allows $H<1/2$, and show in $\S$ 4 of this article that evidence from data points towards medium range dependence, $H<1/2$, in the volatility of the S\&P500 index.

In the setting of (\ref{eq:S})--(\ref{eq:X}), partial observations over $X$ correspond to direct observations
from the price process $S$, that is for times $0<t_1<\cdots < t_n=\ell$, for some $n\ge 1$, we have
\begin{equation}
\label{eq:data}
Y_{k} = \log S_{t_k}\,\,\,(k=1,\ldots,n) \ ,\quad Y=\{Y_{1}, \ldots,Y_n\}\ .
\end{equation}
Given $Y$ we aim at making inference for all parameters $\theta = (\mu,\zeta, H, x_0)$ involved in our  model in (\ref{eq:S})-(\ref{eq:X}).
Available inference methods in this partial observation setting are limited.
\cite{ComRen1998} and \cite{ComEtAl2012} extract information on the spot volatility from the quadratic variation of the price, which is subsequently used to estimate $\theta$. \cite{Ros2008} links the squared increments of the observed price process with the volatility and constructs a wavelet estimator of $H$.
A common feature of these approaches, as of other related ones \citep{GloHof2004}, is that they require high-frequency observations. The method in \cite{ChrVie2012a,ChrVie2012b} operates in principle on data of any frequency and estimates $H$
in a non-likelihood manner by calibrating estimated option prices over a grid of values of $H$ against observed market prices. In this paper we develop a computational framework for performing full principled Bayesian inference based on data augmentation. Our approach is applicable even to low frequency data. Existing consistency results in high-frequency asymptotics about estimates of $H$ in a stochastic volatility setting point to slow convergence rates of estimators of $H$ \citep{Ros2008}. In our case, we rely on the likelihood to retrieve maximal information from the data at hand, so our method could contribute at developing a clear empirical understanding for the amount of such information, strong or weak.

Our algorithm presented in this paper has the following characteristics:
\begin{itemize}
\item[a)] the computational cost per algorithmic step is $\mathcal{O}(N\log N)$;
\item[b)] the algorithmic mixing time is mesh-free, $\mathcal{O}(1)$, with respect to the number of imputed
points $N$. That is, reducing the discretization error will not worsen its convergence properties, since the algorithm is well-defined even when considering the complete infinite-dimensional latent path $X$;
\item[c)] it decouples the full dependence between $X$ and $H$; and
\item[d)] it is based on a version of hybrid Monte Carlo, employing Hamiltonian
dynamics to allow big steps in the state space, while treating big blocks of $X$. In examples the whole of the $X$-path and parameter $\theta$ are updated simultaneously.
 \end{itemize}

Markov chain Monte Carlo methods with mesh-free mixing times for distributions which are change of measures from
Gaussian laws in infinite dimensions have already appeared \citep{cott:13} with the closest references for hybrid Monte Carlo being \cite{besk:11,besk:14}. A main methodological contribution of this work is to assemble a number of techniques, including: the Davies and Harte reparameterization, to re-express  the latent-path part of the posterior as a change of measure from an infinite-dimensional Gaussian law; a version of hybrid Monte Carlo which is particularly effective when run on the contrived infinite-dimensional latent-path space; and a careful joint update for path and parameters, enforcing $\mathcal{O}(N\log N)$ costs for the complete algorithm.

\section{Davies and Harte sampling and reparameterization}
\label{sec:DaviesHarte}
\subsection{Fractional Brownian motion sampling}
Our Monte Carlo algorithm considers the driving fractional noise on a  grid of discrete times. We use the Davies and Harte method, sometimes also called the circulant method, to construct $\{B_t^{H},0\le t\le \ell\}$ on the regular grid $\{\delta, 2\delta, \ldots, N\delta\}$ for some $N\ge 1$ and mesh-size $\delta = \ell/N $. The algorithm samples the grid points via a linear transform from independent standard Gaussians. This transform will be used in $\S$ \ref{sub:re} to decouple the latent variables from the Hurst parameter $H$.
The computational cost is $\mathcal{O}(N\log N)$ due to the use of the fast Fourier transform.
The method is based on the stationarity of the increments of fractional Brownian motion on the regular grid  and, in particular, exploits the Toeplitz structure of the covariance matrix of the increments; see \cite{wood:94} for a complete description.

We briefly present the Davies and Harte method following \cite{wood:94}.
We define the $(2N)\times (2N)$ unitary matrix $P$ with elements
$P_{jk} = (2N)^{-1/2} \, \exp\{ -2\pi i\,jk/(2N)\}$, for $0\le j,k \le 2N-1$,
where $i^2=-1$. Consider also the $(2N)\times (2N)$ matrix
$$
Q = \left(
\begin{array}{cc}
Q_{11} &  Q_{12}  \\
Q_{21} & Q_{22}
\end{array}
\right) \ ,
$$
for the following $N\times N$ sub-matrices: $Q_{11}=\mathrm{diag}\{1, {2^{-1/2}},\ldots, 2^{-1/2}\}$;
$Q_{12} = \{q_{ij}\}$ with $q_{i,i-1}=2^{-1/2}$ if $i=1,\ldots, N-1$, otherwise $q_{ij}=0$;
$Q_{21} = \{q_{ij}\}$ with $q_{i,N-i}=2^{-1/2}$ if $i=1,\ldots, N-1$, otherwise $q_{ij}=0$;
$Q_{22}=\mathrm{diag}_{\mathrm{inv}}\{1, {-i}\,2^{-1/2},{-i}\,2^{-1/2},\ldots, {-i}\,2^{-1/2}\}$, where
$\mathrm{diag}_{\mathrm{inv}}$ denotes a matrix with non-zero entries at the inverse diagonal.
%
We define the diagonal matrix
$
\Lambda_{{H}} =
\mathrm{diag}\{\lambda_0,\lambda_1,\ldots, \lambda_{2N-1}\}$
%
for the values
\begin{equation*}
\lambda_k = \sum_{j=0}^{2N-1}  c_{j} \exp \big(  -2\pi i\,\tfrac{jk}{2N}  \big)  \quad  (k=0,\ldots, 2N-1)\ ,
\end{equation*}
Here, $(c_0, c_1, \ldots, c_{2N-1}) = (g(0), g(1),\ldots, g(N-1), 0, g(N-1),\ldots, g(1))$, where $g(k)$ denotes the auto-covariance of increments of $B^{H}$ of lag $k=0,1,\ldots$, that is
\begin{equation*}
g(k)=\Exp\,\{\,B_1^{H}\,(B_{k+1}^H-B_{k}^H)\,\} = \tfrac{1}{2}|k+1|^{2H} + \tfrac{1}{2}|k-1|^{2H}
- |k|^{2H}\ .
\end{equation*}
The definition of  $c_j$'s implies that the $\lambda_k$'s are all real numbers. The Davies and Harte method for generating $B^{H}$ is shown in Algorithm \ref{alg:sMC}.
Finding $Q\,Z$ costs $\mathcal{O}(N)$. Finding  $\Lambda_H$
and then calculating $P \Lambda_H^{1/2}\,Q\,Z$ costs $\mathcal{O}(N\log N)$ due to a fast Fourier transform. Separate approaches prove that the $\lambda_k$'s are non-negative for any $H\in (0,1)$, thus $\Lambda_H^{1/2}$ is well-posed \citep{crai:03}.
%
%
%
There are several other methods to sample a fractional Brownian motion; see for instance \cite{diek:04}.
However, the Davies and Harte method is, to the best of our knowledge, the fastest exact method on a regular grid and boils down
to a simple linear transform that can be easily differentiated, which is needed by our method.

\begin{algorithm}[!h]
\caption{Simulation of stationary increments $(B_\delta^H, B_{2\delta}^H-B_{\delta}^H,\ldots, B_{N\delta}^{H}-B_{(N-1)\delta}^{H})$.} \label{alg:sMC}
\vspace*{-12pt}
\begin{tabbing}
   \enspace (i) Sample $Z\sim \mathcal{N}(0,I_{2N})$. \\
   \enspace (ii) Calculate $Z'=\delta^H\,P\, \Lambda_{{H}}^{1/2}\,Q\,Z$.\,\, \\
   \enspace (iii) Return the first $N$ elements of $Z'$.
\end{tabbing}
\end{algorithm}

\subsection{Reparameterization}
\label{sub:re}
Algorithm~\ref{alg:sMC} gives rise to a linear mapping
\begin{equation*}
Z \mapsto (B^{H}_{\delta}, \ldots, B^{H}_{N\delta})
\end{equation*}
to generate $B^{H}$ on a regular grid of size $N$ from $2N$ independent standard Gaussian variables. Thus, the latent variable principle described in the $\S$ 1 is implemented using
the vector $Z$, a priori independent from $H$, rather than the solution $X$ of (\ref{eq:big}).
Indeed, we work with the joint posterior of $(Z,\theta)$ which has a density with respect to $\otimes_{i=1}^{2N} \mathcal{N}(0,1)\times \mathrm{Leb}_{q}$, namely the product of $2N$ standard Gaussian laws and the $q$-dimensional Lebesgue measure. Analytically, the posterior distribution $\Pi_N$ for $(Z,\theta)$ is specified as follows
\begin{equation}
\label{eq:target}
\frac{d\Pi_N}{d\{\otimes_{i=1}^{2N} \mathcal{N}(0,1)\times \mathrm{Leb}_{q}\}} (Z,\theta\mid Y) \propto \,p(\theta)\,
p_N(Y\mid Z,\theta)\ .
\end{equation}
Subscript $N$, used in the expressions in (\ref{eq:target}) and in the sequel, emphasizes
the finite-dimensional approximations due to involving an $N$-dimensional proxy for the theoretical infinite-dimensional path $X$. Some care is needed here, as standard Euler schemes might not converge when used to approximate stochastic integrals driven by fractional Brownian motion. We explain this in $\S$ \ref{sec:num} and detail the numerical scheme in the Supplementary Material.
The target density can be written as
\begin{equation}
\label{eq:new}
\Pi_N(Z,\theta) \propto  e^{-\frac{1}{2}\langle Z, Z\rangle - \Phi(Z,\theta)}
\end{equation}
where, in agreement with (\ref{eq:target}), we have defined
\begin{equation}
\label{eq:Phi}
\Phi(Z,\theta) = - \log p(\theta) - \log p_N(Y\mid Z,\theta) \ .
\end{equation}
In $\S$ \ref{sec:advHMC} we describe an efficient Markov chain Monte Carlo sampler tailored to sampling from (\ref{eq:new}).

\subsection{Diffusions driven by fractional Brownian motion and their approximation}
\label{sec:num}
For the stochastic differential equation (\ref{eq:big}) and its non-scalar extensions, there is an extensive literature involving various definitions of stochastic integration with respect to $B^{H}$ and determination of a solution; see \cite{biag:08}. For scalar $B^{H}$, the Doss--Sussmann representation \citep{suss:78}
provides the simplest framework for interpreting (\ref{eq:big}) for all $H\in(0,1)$; see also \cite{lysy:13}. It involves a pathwise approach, whereby for any $t\mapsto B^H_t(\omega)$ one obtains a solution of the differential equation for all continuously differentiable paths in a neighborhood of $B^H_{\cdot}(\omega)$ and considers the value of this mapping at $B^H_{\cdot}(\omega)$. Conveniently, the solution found in this way follows the rules of standard calculus  and coincides with the Stratonovich representation when $H=1/2$.

The numerical solution of a fractional stochastic differential equations is a topic of intense investigation \citep{mish:08}. As shown in \cite{lysy:13}, care is needed, as a standard Euler scheme applied to $B^{H}$-driven multiplicative stochastic integrals might diverge to infinity for $H<1/2$. When allowing $H<1/2$ we must restrict attention to a particular family of models to give a practical method. For the stochastic volatility class in (\ref{eq:S})--(\ref{eq:X}) we can assume a Sussmann solution for the volatility equation (\ref{eq:big}). For the corresponding numerical scheme, one can follow \cite{lysy:13} and use the Lamperti transform
$$
F_t = \int^{X_t}\sigma_X^{-1}(u,\zeta)du
$$
so that $F_t$ has additive noise. A standard Euler scheme for $F_t$ will then converge to the analytical solution in an appropriate mode, under regularity conditions. In principle this approach can be followed for general models with scalar differential equation and driving $B^{H}$. The price process
differential equation (\ref{eq:S}) is then interpreted in the usual It\^o way. In $\S$ \ref{sec:application}, we will extend the model in (\ref{eq:S})--(\ref{eq:X}) to allow for a leverage effect. In that case the likelihood $p(Y\mid B_H)$ will involve a multiplicative stochastic integral over $B_H$. Due to the particular structure of this class of models the integral can be replaced with a Riemannian one, allowing for a standard finite difference approximation scheme. The Supplementary Material details the numerical method used in the applications. For multi-dimensional models one cannot avoid multiplicative stochastic integrals. For $H>1/2$ there is a well-defined framework for the numerical approximation
of multiplicative stochastic integrals driven by $B^{H}$, see \cite{hu:13}. For $1/3<H<1/2$ one can use a Milstein-type scheme, with third order schemes required for $1/4<H\le 1/3$ \citep{deya:12}.

\section{An Efficient Markov chain Monte Carlo Sampler}
\label{sec:advHMC}
\subsection{Standard hybrid Monte Carlo algorithm}
We use the hybrid Monte Carlo algorithm to explore the posterior of $Z,\theta$ in (\ref{eq:new}). The standard method was introduced in \cite{duna:87}, but we employ an advanced version, tailored to the structure of the distributions of interest and closely related to algorithms developed in \cite{besk:11,besk:14} for effective sampling of change of measures from Gaussian laws in infinite dimensions. We first briefly describe the standard algorithm.

The state space is extended via the velocity $v=(v_{z},v_{\theta})\in\mathbb{R}^{2N+q}$.
The original arguments $x=(z,\theta)\in\mathbb{R}^{2N+q}$ can be thought of as location.
The total energy function is, for $\Phi$  in (\ref{eq:Phi}),
\begin{equation}
\label{eq:energy0}
H(x,v;M) =\Phi(x)+\tfrac{1}{2}\langle z, z \rangle +
\tfrac{1}{2} \langle v, M v\rangle \ ,
\end{equation}
for a user-specified positive-definite mass matrix $M$, involving the potential $\Phi(x)+\langle z,z \rangle/2$
and kinetic energy $\langle v, Mv \rangle/2$.
Hamiltonian dynamics on $\R^{2N+q}$ express preservation of energy and are defined via the system of differential equations $dx/dt =M^{-1}(\partial H/\partial v)$,
$M\,(dv/dt)=-\partial H/\partial x$,  %
which in the context (\ref{eq:energy0}) become
$dx/dt=v$,  $M\,(dv/dt)=-(z,0)^{\top} -\nabla \Phi(x)$.
In general, a good choice for $M$ resembles the inverse covariance
of the target $\Pi_N(x)$. In our context, guided by the prior structure of $(z,\theta)$, we set
\begin{equation}
\label{eq:mass}
M = \left(
\begin{array}{cc}
I_{2N} &  0  \\
0 & A
\end{array}
\right) \ , \quad A = \mathrm{diag}\{a_i:i=1,\ldots, q\}\  .
\end{equation}
and rewrite the Hamiltonian equations as
\begin{equation}
\label{eq:hamiltonf}
dx/dt=v\ ,\quad  dv/dt=-(z,0)^{\top} -M^{-1}\,\nabla \Phi(x) \ .
\end{equation}

The standard hybrid Monte Carlo algorithm discretizes
(\ref{eq:hamiltonf}) via a leapfrog scheme, so that for $h>0$:
\begin{align}
v_{h/2} &= v_{0} - \tfrac{h}{2}\,(z_0,0)^{\top}  - \tfrac{h}{2}\,\,M^{-1}\,\nabla
\Phi(x_0)\ , \nonumber \\
x_h &= x_0 + h\,v_{h/2} \ ,\label{eq:lp}\\
v_h &= \,v_{h/2} - \tfrac{h}{2}\,(z_h,0)^{\top}-\tfrac{h}{2}\,\,M^{-1}\,\nabla \Phi(x_h)
\ . \nonumber
\end{align}
Scheme (\ref{eq:lp}) gives rise to the operator
$(x_0, v_0)\mapsto \psi_h(x_0,v_0) = (x_h, v_h)$.
The sampler looks
up to a time horizon $T>0$  via the synthesis of
%
$I=\lfloor T/h \rfloor$
%
leapfrog steps,  so we define $\psiT$ to be the synthesis of $I$ mappings $\psih$.
The dynamics in (\ref{eq:hamiltonf}) preserve the total energy and are invariant for the density
$\exp\{-H(x,v;M)\}$, but their discretized version requires an accept/reject correction.
The full method is shown in Algorithm \ref{alg:sHMC2}, with $\mathcal{P}_x$ being d projection on $x$.
The proof that Algorithm \ref{alg:sHMC2} gives a Markov chain that preserves $\Pi_N(x)$ is
based on $\psiT$ being volume-preserving and having the symmetricity property
$\psih^{I}(x_I,-v_I) =(x_0,-v_0)$, as with the exact solver of the Hamiltonian equations,
see  for example \cite{duna:87}.
\begin{algorithm}[!h]
\caption{Standard hybrid Monte Carlo algorithm, with target  $\Pi_N(x)=\Pi_{N}(Z,\theta)$ in (\ref{eq:new}).} \label{alg:sHMC2}
\vspace*{-12pt}
\begin{tabbing}
   \enspace (i) Start with an initial value $x^{(0)}\in \R^{2N+q}$ and set $k=0$. \\
   \enspace (ii) Given $x^{(k)}$ sample  $v^{(k)} \sim \mathcal{N}(0,M^{-1})$ and propose
$x^{\star}=\mathcal{P}_x\,\psiT(x^{(k)},v^{(k)})$.\\
   \enspace (iii) Calculate  $a= 1\wedge \exp\{ H(x^{(k)},v^{(k)};M)-H(\psi_h ^I(x^{(k)},v^{(k)});M)\}$.\\
   \enspace (iv) Set $x^{(k+1)}=x^{\star}$ with probability $a$; otherwise set
$x^{(k+1)}=x^{(k)}$.\\
   \enspace (v) Set $k \to k+1$ and go to (ii).
\end{tabbing}
\end{algorithm}
\begin{remark}
Index $t$ of Hamiltonian equations must not be confused with index $t$ of
the diffusion processes in the models of interest. When applied here, each hybrid Monte Carlo step updates
a complete sample path, so the $t$-index for paths can be regarded as
a space direction.
\end{remark}
\subsection{Advanced hybrid Monte Carlo algorithm}

Algorithm \ref{alg:sHMC2} provides an inappropriate proposal $x^{\star}$ for increasing
$N$ \citep{besk:11} with the acceptance probability approaching $0$, when $h$ and $T$ are fixed. Indeed,
\cite{besk:13} suggest that controlling the acceptance probability requires step-size $h=\mathcal{O}(N^{-1/4})$.
Advanced hybrid Monte Carlo simulation avoids this degeneracy by employing a modified leapfrog scheme that gives better performance in high dimensions.

\begin{remark}
The choice of mass matrix $M$ as in (\ref{eq:mass}) is critical for the final algorithm. Choosing $I_{2N}$ for the upper-left block of $M$ is motivated by the prior for $Z$. We will see in $\S$\ref{sec:well} that this choice also provides the well-posedness of the algorithm as $N\rightarrow \infty$. A posteriori, we have found empirically that the information in the data spreads fairly uniformly over the many $Z_{i}$ $(i=1,\ldots, 2N)$, thus $I_{2N}$ seems a sensible choice also under this viewpoint. For the choice of the diagonal $A$, in the numerics we have tried to
resemble the inverse of the marginal posterior variances of $\theta$ as estimated by preliminary
runs. More automated choices could involve adaptive Markov chain Monte Carlo  or even recent Riemannian manifold approaches \citep{giro:11} using the Fisher information. We will not go into such directions in the paper as even a less contrived choice of $M$ gives efficient methods.
\end{remark}

\begin{remark}
\label{rem:diff}
The development below is closely related
to the approach in \cite{besk:11}, who illustrate the mesh-free mixing property of the algorithm in the context of distributions of diffusion paths driven by Brownian motion. In this paper, the algorithm is extended to also take under consideration the involved parameters and the different set-up with a product of standard Gaussians as the high-dimensional Gaussian reference measure.
\end{remark}

The method develops as follows. Hamiltonian equations (\ref{eq:hamiltonf}) are  now split into two parts
\begin{align}
\label{eq:s1}
dx/dt&=0\ , &dv/dt&=-M^{-1}\,\nabla\Phi(x)\ ;\\
\label{eq:s2}
dx/dt&=v\ , &dv/dt&=-(z,0)^{\top} \ ,
\end{align}
where both equations can be solved analytically. We obtain a numerical integrator for (\ref{eq:hamiltonf}) by synthesizing the steps of (\ref{eq:s1}) and (\ref{eq:s2}). We define the solution operators of (\ref{eq:s1}) and (\ref{eq:s2})
\begin{align}
\label{eq:xi1} \Xi_t(x,v) &= (x,\,v-t\,M^{-1}\,\nabla\Phi(x))\ ;\\
\label{eq:xi2}
\tilde{\Xi}_t(x,v) &= \big( \,( \cos(t)\,z+\sin(t)\,v_z,\, \theta + t\,v_{\theta}  )\,,
(-\sin(t)\,z+\cos(t)\,v_z, v_{\theta})\,\big)\ .
\end{align}
The numerical integrator for (\ref{eq:hamiltonf}) is defined as
\begin{equation}
\label{eq:psih}
\Psi_h = \Xi_{h/2}\circ\XiL\circ\Xi_{h/2}\ ,
\end{equation}
for small $h>0$. As with the standard hybrid Monte Carlo algorithm, we synthesize $I=\lfloor T/h \rfloor $ leapfrog steps $\Psi_h$ and
denote the complete mapping $\PsiT$.
Notice that $\Psi_h$ is volume-preserving and that, for $(x_h,v_h) = \Psi_{h}(x_0,v_0)$, the symmetricity property
$\Psi_h(x_h,-v_h) =(x_0,-v_0)$ holds.
Due to these  properties, the acceptance probability has the same expression as with the standard hybrid Monte Carlo algorithm.
The full method is shown in Algorithm \ref{alg:aHMC}.

%
%
\begin{algorithm}[!h]
\caption{Advanced hybrid Monte Carlo, with target $\Pi_N(x)=\Pi_N(Z,\theta)$ in (\ref{eq:new}).}\label{alg:aHMC}
\vspace*{-12pt}
\begin{tabbing}
   \enspace (i) Start with an initial value $x^{(0)}\sim \otimes_{i=1}^{2N}\mathcal{N}(0,1)\times p(\theta)$ and set $k=0$. \\
   \enspace (ii) Given $x^{(k)}$ sample  $v^{(k)} \sim \mathcal{N}(0,M^{-1})$ and propose
$x^{\star}=\mathcal{P}_x\,\PsiT(x^{(k)},v^{(k)})\ .$\\
   \enspace (iii) Calculate $a= 1\wedge \exp\{H(x^{(k)},v^{(k)};M) -  H(\PsiT(x^{(k)},v^{(k)});M)\}$. \\
   \enspace (iv) Set $x^{(k+1)}=x^{\star}$ with probability $a$; otherwise set
$x^{(k+1)}=x^{(k)}.$\\
   \enspace (v) Set $k \to k+1$ and go to (ii).
\end{tabbing}
\end{algorithm}
\subsection{Well-Posedness of advanced hybrid Monte Carlo when $N=\infty$.}
\label{sec:well}

An important property for the advanced method is its mesh-free mixing time.
As $N$ increases and $h,T$ are held fixed, the convergence/mixing properties of the Markov chain
do not deteriorate. To illustrate this, we show that there is a well-defined algorithm in the limit $N=\infty$.

\begin{remark}
We follow closely \cite{besk:14},
with the differences in the current set-up discussed in Remark \ref{rem:diff}.
We include a proof of the well-posedness of advanced hybrid Monte Carlo when $N=\infty$ here as it cannot be directly implied from \cite{besk:14}.
The proof provides insight into the algorithm, for instance highlighting the aspects that deliver mesh-free mixing.
\end{remark}

Denote the vector of partial derivatives over the $z$-component with $\nabla_z$,
so that we have $\nabla_x = (\nabla_z,\nabla_\theta)^{\top}$. Here, $z\in \R^{\infty}$,
and the distribution of interest corresponds to
$\Pi_N$ in (\ref{eq:new}) for $N=\infty$, denoted
by $\Pi$ and defined on the infinite-dimensional space
%
$\Hs  = \R^{\infty}\times\R^{q}$
via the change of measure
\begin{align}
\label{eq:targetI}
\frac{d\Pi}{d\{\otimes_{i=1}^{\infty} \mathcal{N}(0,1)\times Leb_{q}\}} &(Z,\theta\mid Y) \propto  e^{-\Phi(Z,\theta)}
\end{align}
for a function $\Phi: \Hs \mapsto \R$. Also,
we need the vector of partial derivatives
$\nabla \Phi: \Hs \mapsto \Hs$. We have the velocity $v=(v_z,v_\theta)\in \Hs$,
whereas the matrix $M$, specified in (\ref{eq:mass})
for finite dimensions, has the infinite-dimensional
identity matrix $I_{\infty}$ at its upper-left block instead of $I_{2N}$.
Accordingly, we have that $\Xi_{h/2}, \XiL, \Psi_h: \Hs\times \Hs \mapsto \Hs\times \Hs$.

We consider the joint location/velocity law on $(x,v)$, $Q(dx,dv)= \Pi(dx)\otimes \mathcal{N}(0,M^{-1})(dv)$.
The main idea is that $\Psi_h$ in (\ref{eq:psih}) projects $(x_0,v_0)\sim Q$ to
$(x_h,v_h)$ having a distribution absolutely continuous
with respect to $Q$, an attribute that implies existence of a non-zero
acceptance probability when $N=\infty$, under conditions on $\nabla \Phi$.
This is apparent for $\XiL$ in (\ref{eq:xi2}) as it applies a rotation in the $(z,v_z)$-space which is invariant for $\prod_{i=1}^{\infty} \mathcal{N}(0,1)\otimes \prod_{i=1}^{\infty} \mathcal{N}(0,1)$; thus the overall step preserves absolute continuity
of $Q(dx,dv)$. Then, for step $\Xi_{h/2}$ in (\ref{eq:xi1}), the gradient $\nabla_z \Phi(z,\theta)$ must lie in the
so-called Cameron--Martin space of $\prod_{i=1}^{\infty}\mathcal{N}(0,1)$ for the translation
$v\mapsto v - (h/2)\,M^{-1}\nabla \Phi(x)$ to preserve absolute continuity of the $v$-marginal $Q(dv)$. This Cameron--Martin
space is that of squared summable infinite vectors $\ell_2$ \citep[Chapter 2]{prat:92}. In contrast, for the
standard hybrid Monte Carlo algorithm one can consider even the
case of $\Phi(x)$ being a constant, so that $\nabla \Phi\equiv 0$, to
see that, immediately from the first step in the leapfrog update in (\ref{eq:lp}), an input sample from the target $Q$
gets projected to a variable with singular law with respect to $Q$ when
$N=\infty$, thus has zero acceptance probability.

For a rigorous result, we first define a reference measure on the $(x,v)$-space:
\begin{equation*}
Q_0 = Q_0(dx,dv) = \Big\{\prod_{i=1}^{\infty}\mathcal{N}(0,1)  \otimes \mathrm{Leb}_q \Big\}(dx)
 \otimes\mathcal{N}(0,M^{-1})(dv)\  ,
\end{equation*}
so that the joint target is
%
$Q(dx,dv) \propto  \exp\{-\Phi(x)\}\,Q_0(dx,dv)$.
%
We also consider the sequence of probability measures on $\Hs\times \Hs$ defined as
%
$Q^{(i)} = Q\circ \PsiInv{i}$ $(i=1,\ldots, I)$
%
corresponding to the push-forward projection of
$Q$ via the leapfrog steps.
For given $(x_0,v_0)$, we write $(x_i,v_i) = \Psi_h^{\,i}(x_0,v_0)$.
The difference in   energy
$\Delta H(x_0,v_0)$ appearing in the statement of Proposition \ref{pr:1} below is still defined
as $\Delta H(x_0,v_0) = H(x_I,v_I;M)-H(x_0,v_0;M)$ for the energy function
in (\ref{eq:energy0}) with the apparent extension of the involved inner product
on $\R^{\infty}$. Even if $H(x_0,v_0;M)=\infty$ with probability 1,
the difference $\Delta H(x_0,v_0)$ does not explode, as implied by the analytic expression for $\Delta H(x_0,v_0)$ given in the proof of Proposition \ref{pr:1} in the Appendix.
\begin{proposition}
\label{pr:1}
Assume that $\nabla_{z}\Phi(z,\theta)\in \ell_2$, almost surely under
$\prod_{i=1}^{\infty}\mathcal{N}(0,1)\otimes p(d\theta)$. Then:
\begin{itemize}
\item[i)] $Q^{(I)}$ is absolutely continuous with respect to $Q_0$ with
probability density,
\begin{equation*}
\frac{dQ^{(I)}}{dQ_0}(x_I,v_I) = \exp\{ \Delta H(x_0,v_0) - \Phi(x_I) \}\ .
\end{equation*}
\item[ii)] The Markov chain with transition dynamics, for current position
$x_0\in \Hs$,
\begin{equation*}
x'  = \mathrm{I}\,\{\,U\le a(x_0,v_0)\,\}\,x_{I} + \mathrm{I}\,\{\,U>a(x_0,v_0)\,\}\,x_0\ ,
\end{equation*}
for $U\sim \mathrm{Un}\,[0,1]$ and noise
$v_0\sim \prod_{i=1}^{\infty}\mathcal{N}(0,1)\otimes \mathcal{N}_q(0,A^{-1})$,
has invariant distribution $\Pi(dx)$ in (\ref{eq:targetI}).
\end{itemize}
\end{proposition}

The proof is given in the Appendix.
\begin{remark}
Condition $\nabla_z \Phi(z,\theta)\in \ell_2$ relates with the fact that the data have a finite amount
of information about $Z$,
so the sensitivity of the likelihood for each individual $Z_i$ can be small for large $N$.
We have not pursued an analytical investigation of this,
as Proposition~\ref{pr:1} already highlights the structurally important  mesh-free property of the method.
\end{remark}

\section{Fractional stochastic volatility models}
\label{sec:application}
\subsection{Data and model}
To illustrate the algorithm, we return to the  fractional stochastic volatility models in (\ref{eq:S})--(\ref{eq:X}). Starting from (\ref{eq:S})--(\ref{eq:X}), we henceforth work with $U_t=\log(S_t)$ and use It\^o's formula to rewrite the equations in terms of $U_t,X_t$. Also, we extend the model by allowing $W_t$ and $B_t^H$ to be correllated
\begin{align}
    dU_t&=(\mu -\sigma_S(X_t)^2/2)\,dt + \sigma_S(X_t)\big\{(1-\rho^2)^{1/2}\,dW_t +\rho\,dB^H_t\big\}\ ,
\nonumber \\
    dX_t&=b_X(X_t,\zeta) dt+\sigma_X(X_t,\zeta) dB_t^H\ ,\quad 0\le t\le \ell\ , \label{eq:model0}
\end{align}
for a parameter $\rho\in (-1,1)$, so henceforth $\theta = (\mu,\zeta, H, \rho, x_0)\in \R^{q}$ with $q=p+4$. We set $H\in (0,1)$, thus allowing for medium range dependence, as opposed to previous literature which typically restricts attention to $H\in (1/2,1)$.
Given the observations $Y$ from the log-price process in (\ref{eq:data}) there is a well-defined likelihood
$p(Y\mid B^{H},\theta)$. Conditionally on the latent driving noise $B^{H}$, the log-price process $U$ is
Markovian. From the specification of the model, we
have that
\begin{equation}
\label{eq:conditional}
  Y_k\mid Y_{k-1},B^{H},\theta\;\sim\; \mathcal{N}\{m_k(B^{H},\theta),\Sigma_k(B^{H},\theta)\}\,k=1,\ldots,n
\end{equation}
where $Y_0\equiv U_0$ assumed fixed,
with mean and variance parameters
\begin{align*}
m_k(B^{H},\theta) &= Y_{k-1}+\int_{t_{k-1}}^{t_k}(\mu -\sigma_S(X_t)^2/2) dt + \rho \int_{t_{k-1}}^{t_k}  \sigma_S(X_t)dB_t^{H}\  ;\\
\Sigma_k(B^{H},\theta) &= (1-\rho^2)\int_{t_{k-1}}^{t_k}\sigma_S(X_t)^2dt\ .
\end{align*}
From (\ref{eq:conditional}), it is trivial to write down the complete expression for the likelihood $p(Y\mid B^{H},\theta)$.

Recalling the mapping
$Z  \mapsto (B^{H}_{\delta},B^{H}_{2\delta},\ldots, B^{H}_{N\delta})$
%
from Davies and Harte method in $\S$ \ref{sec:DaviesHarte}, for $N\ge 1$ and discretization step $\delta=\ell/N$, the expression for $p(Y\mid B^{H},\theta)$ in continuous time will provide an expression for $p_N(Y\mid Z,\theta)$ in discrete time upon consideration of a numerical scheme. In $\S$ \ref{sec:num}, we described the Doss--Sussmann interpretation of the stochastic volatility model, which allows a standard finite-difference scheme.
Expressions for $p_{N}(Y\mid Z,\theta)$ and the derivatives $\nabla_{Z}\log p_N(Y\mid Z,\theta)$, $\nabla_{\theta} \log p_{N}(Y\mid Z,\theta)$ required by the Hamiltonian methods are provided in the Supplementary Material.

A strength of our methodology is the ability to handle different types of data from different sources. To illustrate this, we analyse two extended sets of data in addition to observations of $U_t$. The first extension considers volatility proxies, constructed from option prices, as direct observations on $X_t$, as in \cite{sah:kim2009}, \cite{Jones2003} and \cite{str:bog11}. \cite{sah:kim2009} use two proxies from the VIX index. First, they consider a simple unadjusted proxy that uses VIX to directly obtain $\sigma_S(X_t)$ and therefore $X_t$. Second, an adjusted integrated volatility proxy is considered, assuming that the pricing measure has a linear drift; see \S 5.1 \cite{sah:kim2009}. The integrated volatility proxy is also used by \cite{Jones2003} and \cite{str:bog11} to provide observations of $\sigma_S(X_t)$, where additional measurement error is incorporated in the model. We take the simpler approach and use the unadjusted volatility proxy as a noisy measurement device for $\sigma_S(X_t)$, for two reasons. First, our focus is mainly on exploring the behaviour of our algorithm on a different observation regime, so we want to avoid additional subject-specific considerations, such as assumptions on the pricing measure. Second, the difference between the two approaches is often negligible and can possibly be omitted or left to the error term; see for example the simulation experiments in \cite{sah:kim2009} for the Heston model. The approaches of \cite{Jones2003} and \cite{str:bog11} can still be incorporated in our framework. More generally, the problem of combining option and asset prices must be investigated further even in the context of standard Brownian motion.

Following the above discussion, the additional noisy observations from VIX proxies are denoted by $Y^{x}_k$ and are assumed to provide information on $X_{t_k}$ via
\begin{equation}
\label{eq:VIX}
Y^{x}_k=X_{t_k}+\epsilon_k \quad (k=1,\ldots,n) \ ,
 \end{equation}
where $\epsilon_k$ are independent samples from $\mathcal{N}(0,\tau^2)$. We refer as type A to the dataset consisting of observations $Y$ and as type B to the dataset consisting of $Y$ and $Y^{x}$. The second extension builds up on the type B dataset and incorporates intraday observations on $Y$, thus considering two observation frequency regimes; this is referred to as type C.

The parameter $\tau$ controls the weight placed on the volatility proxies in order to form a weighted averaged volatility measurement combining information from asset and option prices. Hence we treat $\tau$ as a user-specified parameter. In the following numerical examples, we set $\tau=0.05$ based on estimates from a preliminary run of the full model to the S\&P500/VIX time series. In the Supplementary Material we give $p_{N}(Y\mid Z,\theta)$, $\nabla_{Z}\log p_N(Y\mid Z,\theta)$, $\nabla_{\theta} \log p_{N}(Y\mid Z,\theta)$ only for the type A case, as including the terms due to the extra data in (\ref{eq:VIX}) is trivial.

\subsection{Illustration on simulated data}
\label{sec:sim}
We apply our method to the model of \cite{ComRen1998}, also used in \cite {ChrVie2012a, ChrVie2012b}, but we also make an extension for correlated noises as in (\ref{eq:model0}), that is we have
\begin{align}
    dU_t&=(\mu -\sigma_S(X_t)^2/2)\,dt + \exp(X_t/2)\big\{(1-\rho^2)^{1/2}\,dW_t +\rho\,dB^H_t\big\}\ ,
\nonumber \\
    dX_t&=\kappa (\mu_X-X_t) dt+\sigma_X dB_t^H\ .\label{eq:model}
\end{align}
The model is completed with priors similar to related literature, such as \cite{Chib2006}. The prior for $\mu_X$ is normal with $95\%$ credible interval spanning from the minimum to the maximum simulated volatility values, or the real VIX observations when these are used, over the entire period under consideration. The prior for $\sigma_X^2$ is an inverse gamma with shape and scale parameters $\alpha = 2$ and $\beta = \alpha \times 0.03 \times 252^{1/2}$. Vague priors are chosen for the remaining parameters; uniforms on $(0,1)$ and $(-1,1)$ for $H$ and $\rho$ and $\mathcal{N}(0,10^6)$ for $\mu$.

We first apply Algorithm \ref{alg:aHMC} to simulated data. We generated 250 observations from model (\ref{eq:model}), corresponding roughly to a year of data. We considered two datasets: i) Sim-A: with 250 daily observations on $S_t$ only, as in (\ref{eq:data}); and ii) Sim-B: with additional daily observations on $X_t$ for the same time period, contaminated with measurement errors as in (\ref{eq:VIX}). We consider $H=0.3$, $0.5$ and $0.7$, and use a discretization step $\delta=0.1$ for the Euler approximation of the path of $Z$, resulting in $2N = 2\times 250 \times 10= 5000$. The true values of the parameters were chosen to be similar to those in previous analyses on the S\&P500/VIX indices based on standard Markovian models \citep{sah:kim2009,Chib2006} and with the ones we found from the real-data analysis in $\S$ \ref{sec:data}. The Hamiltonian integration horizon was set to $T=0.9$ and $T=1.5$ for datasets Sim-A and Sim-B respectively. The number of leapfrog steps was tuned to achieve an average acceptance rate between $70\%$ and $80\%$. Various values between $10$ to $50$ leapfrog steps across the different simulated datasets were used to achieve this.

Traceplots for the case $H=0.3$ are shown in the supplementary materials. We did not notice substantial difference for $H=0.5$ and $H=0.7$, so we do not show the related plots. The mixing of the chain appears to be quite good considering the complexity of the model. Table \ref{tab:SimAsset}  shows posterior estimates obtained from running advanced hybrid Monte Carlo algorithm for datasets Sim-A, Sim-B.

\begin{figure}[!h]
\vspace{-4cm}
\begin{center}
\includegraphics[trim = 1cm 10.5cm 0cm 5cm,clip=true,scale=0.8]{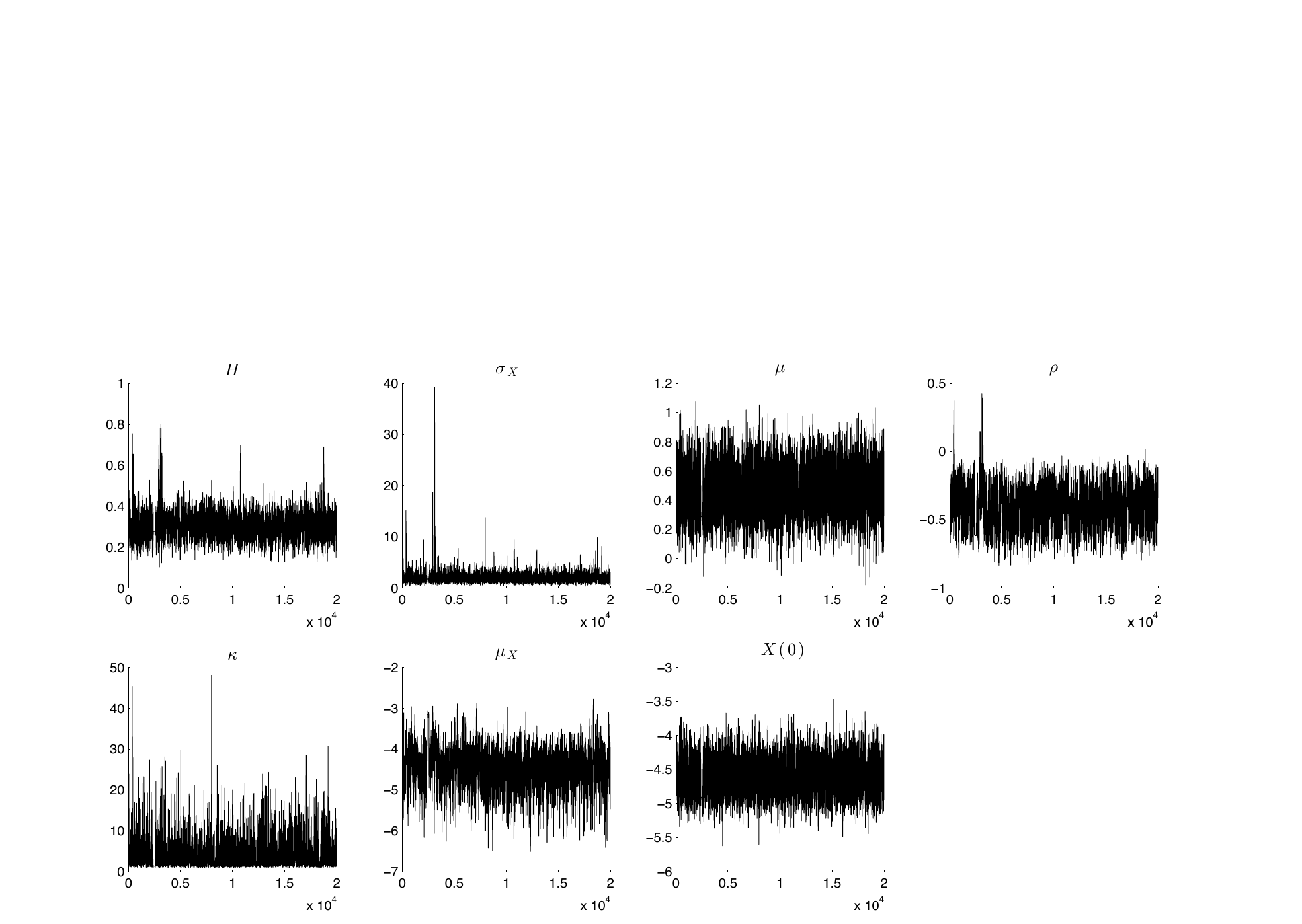}
\end{center}
\vspace{-0.1cm}
\caption{Traceplots  from $2\times 10^4$ iterations of advanced hybrid Monte Carlo,
for dataset Sim-A. True parameter  values are as in Table \ref{tab:SimAsset}
with  $H=0.3$.
Execution time was about 5h, with code in Matlab.}
\label{fig:trace_i}
\end{figure}
\begin{figure}[!h]
\vspace{-4cm}
\begin{center}
\includegraphics[trim = 1cm 10.5cm 0cm 5cm,clip=true,scale=0.8]{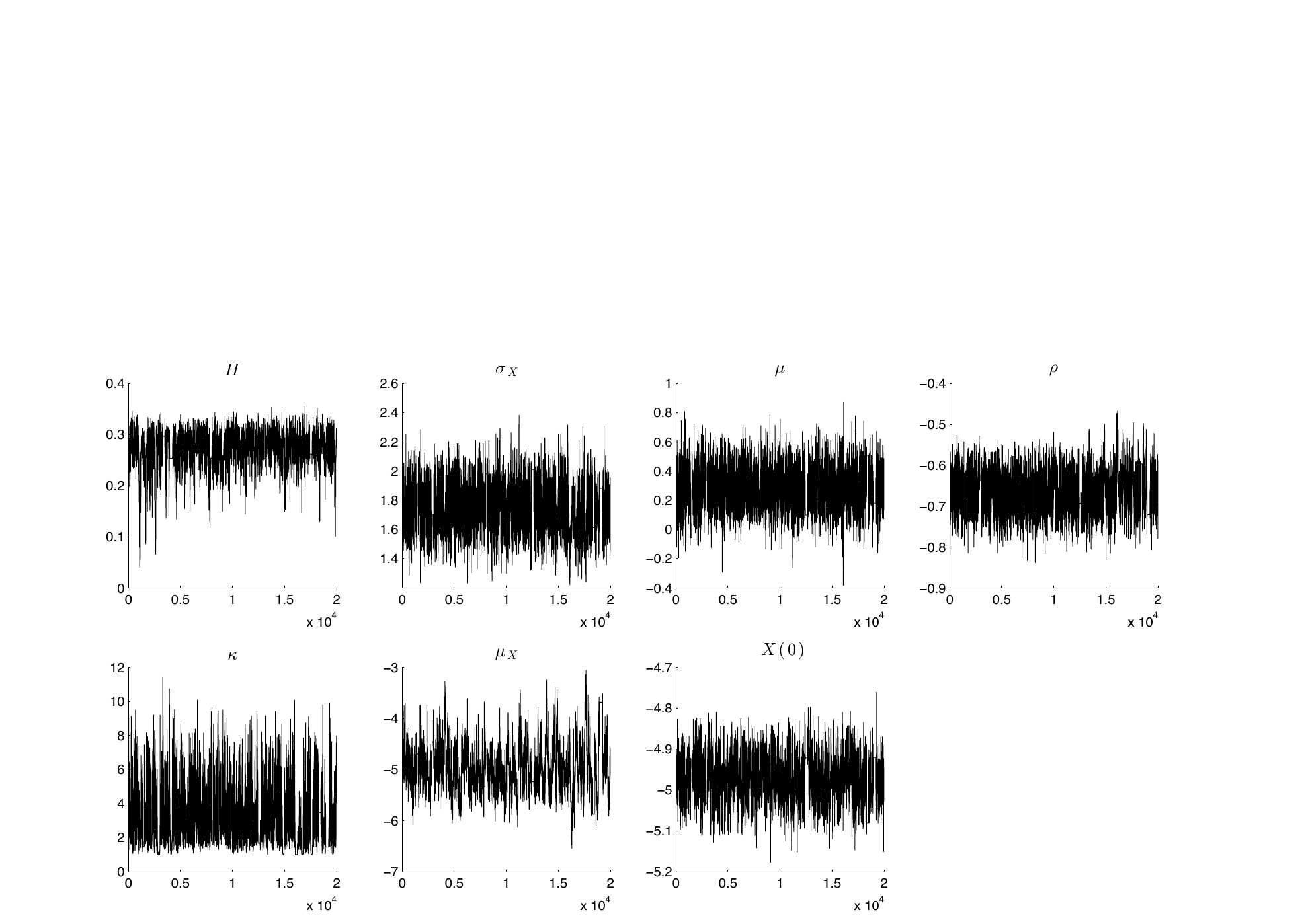}
\end{center}
\vspace{-0.1cm}
\caption{Traceplots as in Figure \ref{fig:trace_i}, for dataset Sim-B,
with true parameter  values as in Table \ref{tab:SimAsset} and $H=0.3$.
Execution time was about 7h, due to using 50 leapfrog steps, whereas
the algorithm for Sim-A used 30. }
\label{fig:trace_ii}
\end{figure}

The results dataset Sim-A in Table \ref{tab:SimAsset} show reasonable agreement between the posterior distribution and the true parameter values. More interestingly, several of  the credible intervals are relatively wide, reflecting the limited amount of data or the small amount of information in Sim-A for particular parameters. Nevertheless, in the case of medium range memory with $H<1/2$,  the 95\% credible interval is below $0.5$; i.e. $[0.201,0.437]$. When $H=0.5$ or $H=0.7$ the credible intervals for $H$ are wider. In particular for $H=0.7$ this may suggest that the data do not provide substantial evidence towards long memory. In such cases one option is to consider richer datasets such as Sim-B where, as can be seen from Table \ref{tab:SimAsset}, the credible interval is tighter and does not contain $0.5$. Another option, not using on volatility proxies, is to consider a longer or a more frequently observed time series using intraday data. For example, re-running the algorithm on a more dense version of the Sim-A dataset containing two equispaced observations per day, yields the 95\% credible interval for $H$ which is $[0.584, 0.744]$. The posterior distribution for Sim-B is more informative for all parameters and provide accurate estimates of $H$. More specifically the 95\% credible interal for $H$ is below $0.5$ when $H=0.3$ and above $0.5$ when $H0.7$.  
\begin{table}[!h]
\def~{\hphantom{0}}
\tbl{Posterior summaries}{%
 \centering
 \begin{tabular}{ c c r r r r r r r r r }
 & & &\multicolumn{4}{c}{{Dataset Sim-A}}&\multicolumn{4}{c}{{Dataset Sim-B}}\\ [5pt]
{Dataset}&{Parameter}&{True value}&{2.5\%}& {97.5\%}&{Mean}&{Median}&{2.5\%}& {97.5\%}&{Mean}&{Median}\\[5pt]
H=0.3 			&$\mu$   		& 0.25    	& 0.18	& 0.76	& 0.46	&0.46        & 0.01	&0.55	&0.28	&0.28 \\
						&$\rho$   		& $-$0.75	& -0.69	&-0.12	&-0.40	&-0.40      	& $-$0.75	&$-$0.57	&$-$0.67	&$-$0.67 \\
						&$\kappa$   	& 4.00	& 1.13	&12.15	&3.79	&2.79        & 1.01	&7.40	&3.22	&2.74 \\
						&$\mu_X$   	& $-$5.00	& $-$5.62	&$-$3.44	&$-$4.46	&$-$4.42  & $-$5.85	&$-$3.74	&$-$4.95	&$-$4.98 \\
						&$H$   		& 0.30 	&0.20	&0.44	&0.30	&0.30        &0.18	&0.32	&0.27	&0.28 \\
						&$\sigma_X$ 	& 2.00	& 0.90	&3.90	&1.95	&1.78        & 1.45	&2.07	&1.75	&1.75\\
						&$X_0$       	& $-$5.00	& $-$5.05	&$-$4.07	&$-$4.59	&$-$4.60  & $-$5.08	&$-$4.87	&$-$4.97	&$-$4.97\\
H=0.5 			&$\mu$   		& 0.25	& 0.01	&0.99	&0.48	&0.470      & $-$0.14	&0.39	&0.14	&0.15 \\
						&$\rho$   		& $-$0.75	& $-$0.91	&$-$0.13	&$-$0.60	&$-$0.62  & $-$0.88	&$-$0.75	&$-$0.82	&$-$0.82\\
						&$\kappa$   	& 4.00	& 1.33	&19.94	&7.38	&6.24        & 2.49	&6.53	&3.96	&3.75\\
						&$\mu_X$   	& $-$5.00	& $-$5.41	&$-$3.94	&$-$4.83	&$-$4.90  & $-$5.90	&$-$3.87	&$-$4.71	&$-$4.61 \\
						&$H$   		& 0.50	& 0.29	&0.74	&0.50	&0.49        & 0.48	&0.55	&0.52	&0.52 \\
						&$\sigma_X$ 	& 2.00	& 0.83	&4.60	&2.29	&2.14        & 1.74	&2.53	&2.10	&2.09\\
						&$X_0$       	& $-$5.00	&$-$5.75	&$-$4.56	&$-$5.15	&$-$5.13  &$-$5.04	&$-$4.87	&$-$4.96	&$-$4.96\\	
H=0.7 			&$\mu$   		& 0.25	& 0.19	&0.38	&0.28	&0.28        & $-$0.09	&0.39	&0.15	&0.14 \\
						&$\rho$   		& $-$0.75	& $-$0.78	&$-$0.25	&$-$0.60	&$-$0.62  & $-$0.79	&$-$0.68	&$-$0.72	&$-$0.73 \\
						&$\kappa$   	& 4.00	& 1.13	&12.12	&4.89	&4.31        & 2.18	&15.57	&6.82	&7.97\\
						&$\mu_X$   	& $-$5.00	& $-$5.65	&$-$4.93	&$-$5.38	&$-$5.42  & $-$5.52	& $-$4.38	&$-$5.02	&$-$5.00\\
						&$H$   		& 0.70	&0.47	&0.80	&0.61	&0.59        &0.62	&0.83	&0.74	&0.73 \\
						&$\sigma_X$   & 2.00	& 0.90	&3.15	&1.72	&1.61        & 1.22	& 5.33	& 2.92	&3.04 \\
						&$X_0$      	& $-$5.00	& $-$5.47	&$-$4.88	&$-$5.07	&$-$5.03  & $-$5.15	&$-$4.97	&$-$5.06	&$-$5.06 \\
\end{tabular}}\vskip8pt
\label{tab:SimAsset}
\end{table}

\subsection{Real data from S\&P500 and VIX time series}
\label{sec:data}

We consider the following datasets:
\begin{itemize}
\item[i)] dataset A, of S\&P500 values only, that is discrete-time observations of the price process. We considered daily S\&P500 values from 5 March 2007 to 5 March 2008, before the Bear Stearns closure, and from 15 September 2008 to 15 September 2009, after the Lehman Brothers bankruptcy,
\item[ii)] dataset B, as above, but with daily VIX values for the same periods added,
\item[iii)] dataset C, as dataset B, but with intraday observations of  S\&P500 obtained from TickData added. For each day we extracted 3 equi-spaced observations from 8:30 to 15:00.
\end{itemize}
Table \ref{tab:data} show posterior estimates  from our algorithm for datasets A, B, C. The integration horizon $T$ was set to $0.9$, $1.5$ and $1.5$ for datasets A, B and C respectively and the numbers of leapfrog steps were chosen to achieve  acceptance probabilities between $70\%$ and $80\%$.

The purpose of this analysis was primarily to illustrate the algorithm in various observation regimes, so we do not attempt to draw strong conclusions from the results. Both extensions of the fractional stochastic volatility model considered in this paper, allowing $H<0.5$ and $\rho\neq 0$, seem to provide useful additions. In all cases the concentration of the posterior distribution of $H$ below $0.5$ suggests medium range dependence, in agreement with the results of \cite{Gath:2014} in high frequency data settings.
Moreover, the value of $\rho$ is negative in all cases, suggesting the presence of a leverage effect. Although parameter estimates are close across the various time periods, types of datasets and time scales, differences may occur in other segments of the S\&P500 data that can shed light in the dynamics of the process and the data. The modeling and inferential framework developed in this paper provide a useful tool for further investigation.

\begin{table}
\def~{\hphantom{0}}
\tbl{Posterior summaries}{
 \centering
 \begin{tabular}{ c c c r r r r r r r }
 & & &\multicolumn{7}{c}{{Parameters}}\\ [5pt]
{}&{} &{} &{$\mu$  }& {$\rho$ }&{$\kappa$}&{$\mu_X$} &{$H$ }& {$\sigma_X$}&{$X_0$}\\[5pt]
dataset - A & 05/03/07 - 05/03/08 		&2.5\%		&$-$0.28	&$-$0.77	&3.33	&$-$5.31	& 0.13	& 0.39	& $-$6.07\\
	         	  & before Bear		 		&97.5\%		&0.19	&$-$0.13	&60.37	&$-$4.27	& 0.40	& 1.42	& $-$5.37\\
		  & Stearns closure 		&Mean		&$-$0.02	&$-$0.47	&26.03	&$-$4.79	& 0.30	& 0.75	& $-$5.71\\
		  & 				 		&Median		&$-$0.01	&$-$0.48	&24.67	&$-$4.78	& 0.31	& 0.70	& $-$5.71\\[5pt]
		  & 15/09/08 - 15/09/09 		&2.5\%		&$-$0.26	& $-$0.73	&1.06	&$-$4.34	& 0.17	& 0.72	& $-$4.84\\
		  & after Lehman 	 		&97.5\%		&0.47	&$-$0.19	&27.26	&$-$2.93	& 0.46	& 3.56	& $-$3.94\\
		  & Brothers closure 			&Mean		&0.10	&$-$0.49	&8.07	&$-$3.63	& 0.36	& 1.61	& $-$4.39\\
		  & 				 		&Median		&0.10	&$-$0.49	&5.83	&$-$3.61	& 0.38	& 1.42	& $-$4.39\\[5pt]
dataset - B & 05/03/07 - 05/03/08 		&2.5\%		&$-$0.12	&$-$0.75	&1.81	&$-$5.28	& 0.25	& 0.59	& $-$5.84\\
	         	  & before Bear		 		&97.5\%		&0.27	&$-$0.50	&7.46	&$-$4.44	& 0.33	& 0.90	& $-$5.65\\
		  & Stearns	closure 		&Mean		&0.07	&$-$0.62	&4.47	&$-$4.93	& 0.29	& 0.72	& $-$5.74\\
		  & 				 		&Median		&0.03	&$-$0.62	&4.47	&$-$4.95	& 0.29	& 0.72	& $-$5.74\\[5pt]
		  & 15/09/08 - 15/09/09 		&2.5\%		&0.08	&$-$0.48	&1.01	&$-$4.50	& 0.34	& 0.60	& $-$4.26\\
		  & after Lehman 	 		&97.5\%		&0.23	&$-$0.19	&2.13	&$-$2.93	& 0.42	& 0.84	& $-$4.08\\
		  & Brothers closure 			&Mean		&0.08	&$-$0.49	&1.35	&$-$3.63	& 0.38	& 0.71	& $-$4.17\\
		  & 				 		&Median		&0.08	&$-$0.49	&1.27	&$-$3.61	& 0.38	& 0.71	& $-$4.17\\[5pt]
dataset - C & 05/03/07 - 05/03/08 		&2.5\%		&$-$0.14	& $-$0.56	&1.10	&$-$5.54 	& 0.26	& 0.68 	&$-$5.80\\
		  & before Bear		 		&97.5\%		&0.32	&$-$0.27	&4.14 	&$-$4.61	& 0.35	& 0.93	& $-$5.61\\
		  & Stearns	closure 			&Mean		&0.10	&$-$0.42	&2.07	&$-$5.07	& 0.31	& 0.80	& $-$5.71\\
		  & 				 		&Median		&0.10	&$-$0.43	&1.86	&$-$5.10	& 0.32	& 0.81	& $-$5.71\\
		  & 15/09/08 - 15/09/09 		&2.5\%		&$-$0.59	& $-$0.48	& 1.21	&$-$4.11	& 0.28	& 0.45	& $-$4.32\\
		  & after Lehman 	 		&97.5\%		&$-$0.33	&$-$0.30	&2.31 	&$-$3.37	& 0.37	& 0.74	& $-$4.08\\
		  & Brothers closure 			&Mean		&$-$0.47	&$-$0.39	&1.54 	&$-$3.75	& 0.33	& 0.59	& $-$4.20\\
		  & 				 		&Median		&$-$0.47	&$-$0.39	&1.46	&$-$3.74	& 0.33	& 0.58	& $-$4.21\\
\end{tabular}}\vskip4pt
\label{tab:dataA}
\end{table}

\subsection{Comparison of different hybrid Monte Carlo implementations}
\label{ssec:NumComparison}

The results in $\S$ \ref{sec:sim} and $\S$ \ref{sec:data} were obtained by updating jointly the latent path and parameters with our method in Algorithm \ref{alg:aHMC}, labelled aHMC$^{joint}$ in the tables that follow. This section contains a quantitative comparison of the performance of aHMC$^{joint}$ against its Gibbs counterpart, aHMC$^{g}$, in which paths and parameters are updated in sequence, and against standard hybrid Monte Carlo in Algorithm \ref{alg:sHMC2}, labelled HMC$^{joint}$, that also jointly updates paths and parameters. In each case, the same mass matrix is used, of the form (\ref{eq:mass}).  We proceed by fixing the integration horizon to $T=0.9$ and $T=1.5$ for datasets Sim-A and Sim-B respectively and the acceptance probability between $70\%$ and $80 \%$, based on previous experience.

Results are summarized in Table \ref{Tab:ESS_dataA}. One way to assess performance  is via the Effective Sample Size (ESS), computed as in \cite{geye:92} from the lagged autocorrelations of the traceplots. ESS provides a measure for the mixing and sampling efficiency of  algorithms, linking to the percentage out of the total number of Monte Carlo draws that can be considered as independent samples from the posterior. We focus on the minimum ESS over the different components of $\theta$ and $Z$, denoted $\min_{\theta}(ESS)$, $\min_{z}(ESS)$ respectively in the tables, with $\min_{\theta,z}(ESS)$ being the overall minimum. Algorithms aHMC$^{joint}$, aHMC$^{g}$, HMC$^{joint}$ were ran on
the datasets Sim-A, Sim-B  with $H=0.3$. Initially the time discretization step of the differential equations was set to $\delta=0.1$ but we also used $\delta=0.01$ for   aHMC$^{joint}$ and HMC$^{joint}$ to illustrate their behaviour as the resolution gets finer. We denote by aHMC$_{\delta=.01}^{joint}$ and HMC$_{\delta=.01}^{joint}$ the algorithms for $\delta=0.01$, with the subscript being omitted for $\delta=0.1$.  The computing time per iteration is  recorded in the  column titled time in the tables and is taken into account  when comparing algorithms.

First, the sampling efficiency over $\theta$ is lower than the one over $Z$ in all cases.
We then compare aHMC$^{joint}$ and aHMC$^{g}$ in both datasets Sim-A and Sim-B.
The joint version is respectively $9.98$ and $5.32$ times more efficient than its Gibbs counterpart, illustrating the effect of a strong posterior dependence between $Z$ and $\theta$. This dependence is introduced by the data since $Z$ and $\theta$ are a-priori independent by construction. These simulations also illustrate the gain provided by the advanced implementation of the hybrid Monte Carlo algorithm over its standard counterpart. In line with the associated theory, this gain increases as the discretization step $\delta$ becomes smaller, resulting into roughly four times more efficient algorithms for $\delta=0.01$.

\begin{table}[!h]
\def~{\hphantom{0}}
\tbl{Relative efficiency of different versions of
hybrid Monte Carlo}{
\begin{tabular}{cccccccc}
			& Sampler 		& $\min_{\theta}(ESS)$ & $\min_{z}(ESS)$&leapfrogs & time & $\frac{\min_{\theta,z}(ESS)}{time}$ & rel.~$\frac{\min_{\theta,z}(ESS)}{time}$\\[5pt]
Dataset Sim-A & aHMC$^{joint}$ & 1.47\% 		   & 3.95\% 	      & 	10	&0.87  & 1.70 								 & 9.98\\
			& aHMC$^{gibbs}$  & 0.15\% 		& 4.05\% 		    &		10	& 0.88 & 0.17 								 & 1.00\\
			& HMC$^{joint}$	 & 1.15\% 			& 1.2\% 		   &		10	& 0.88 & 1.33 								 & 7.81\\
			& aHMC$_{\delta=.01}^{joint}$ & 1.48\% & 4.35\% 	      & 	10	&1.27 & 1.17 								 &4.39 \\
			& HMC$_{\delta=.01}^{joint}$ & 1.35\%   & 3.50\% 	      & 	40	&5.06  & 0.27 								 & 1.00\\
Dataset Sim-B	& aHMC$^{joint}$ & 3.19\% & 8.81\% & 50 & 3.35 & 0.95 & 5.32\\
			& aHMC$^{gibbs}$  & 0.60\% & 5.00\% & 50 &3.41 & 0.18 & 1.00\\
			& HMC$^{joint}$    & 1.2\% & 3.40\%   &  50& 3.35 & 0.36 & 2.00\\
			& aHMC$_{\delta=.01}^{joint}$ & 1.94\% & 8.40\% 	      & 	50	&6.13 & 0.32	& 3.76\\
			& HMC$_{\delta=.01}^{joint}$ & 1.03\%   & 6.95\% 	      & 	100	&12.26  & 0.08 	& 1\\ 
\end{tabular}}
\label{Tab:ESS_dataA}
\begin{tabnote} The algorithms are applied on dataset Sim-A and Sim-B. Comparison is made via the minimum effective sample size and computing times in seconds.
\end{tabnote}
\end{table}


\section{Discussion}
\label{sec:discuss}

Our methodology performs reasonably well and provides, to our knowledge, one of the few options for routine Bayesian likelihood-based estimation for partially observed diffusions driven by fractional noise. Current computational capabilities together with algorithmic improvements allow practitioners to experiment with non-Markovian model structures of the class considered in this paper in generic non-linear contexts.

It is of interest to investigate the implications of the fractional model in option pricing for $H<0.5$. The joint estimation of physical and pricing measures based on asset and option prices can be studied in more depth, both for white and fractional noise. Moreover, the samples from the joint posterior of $H$ and the other model parameters can be used to incorporate parameter uncertainty to the option pricing procedure. The posterior samples can also be used for Bayesian hypothesis testing, although this task may require the marginal likelihood. Also, models with time-varying $H$ are worth investigating when considering long
time series. The Davies and Harte method, applied on blocks of periods of constant $H$ given a stream of standard normals, would typically create discontinuities in conditional likelihoods, so a different and sequential method could turn out to be more appropriate in this context.

Another direction of investigation involves combining the algorithm in this paper, focusing on computational robustness in high dimensions, with recent Riemannian manifold methods \citep{giro:11}
that automate the specification of the mass matrix and perform efficient Hamiltonian transitions on distributions with highly irregular contour structure.

Considering general Gaussian processes beyond fractional Brownian motion,
our methodology can also be applied for models when the latent variables correspond to general
stationary Gaussian processes, as the initial Davies and Harte transform and all other steps in the development of our method can be carried forward in this context. For instance, Gaussian prior models for infinite-dimensional spatial processes is a potential area of application.

We assumed existence of a non-trivial Lebesgue density for observations given the latent
diffusion path and parameters. This is not the case when data correspond to direct observations of the process,
where one needs to work with Girsanov densities for diffusion bridges.
\cite{lysy:13} look at this set-up.

Finally, another application can involve  parametric inference for generalized Langevin equations with fractional noise, with such models arising in physics and biology \citep{kou:04}.

\section*{Acknowledgements}
The second and third authors were supported by an EPSRC grant.
We thank the reviewers for suggestions that greatly improved  the paper.

\vspace{-0.2cm}
\appendix

\section*{Supplementary material}
\label{SM}
Supplementary material available online give the likelihood
$p_{N}(Y\mid Z,\theta)$ and  derivatives
$\nabla_Z p_N(Y\mid Z,\theta)$, $\nabla_{\theta} \log p_{N}(Y\mid Z,\theta)$, required by the Hamiltonian methods
for the stochastic volatility class of models in (\ref{eq:model0}) under the observation regime (\ref{eq:data}).

\appendixone
\section*{Appendix}

\subsection{Proof of Proposition \ref{pr:1}.}

The proof that standard hybrid Monte Carlo  preserves  $Q_{N}(x,v)=\exp\{-H(x,v;M)\}$, with $H$ in (\ref{eq:energy0}), is based on the volume preservation
of $\psi_h^{I}$. That is,  for reference measure $Q_{N,0}\equiv \mathrm{Leb}_{4N+2q}$ we have
 $Q_{N,0}\circ \psi_h^{-I}\equiv Q_{N,0}$, allowing for simple change of variables
when integrating  \citep{duna:87}. In infinite dimensions,
a similar equality for $Q_0$ does not hold, so instead we adopt a probabilistic approach.
To prove (i),  we obtain a recursive formula for the
densities $dQ^{(i)}/dQ_0$ for $i=1,\ldots, I$.
We set
\begin{equation*}
\label{eq:massI}
\Cov = M^{-1} =  \left(
\begin{array}{cc}
I_{\infty} &  0  \\
0 & A^{-1}
\end{array}
\right) \ ,
\end{equation*}
with $A = \mathrm{diag}\{a_1,\ldots,a_q\}$. We also set
%
$g(x) = -\Cov^{1/2}\,\nabla \Phi(x)$, $x\in \Hs$. 
%
From the definition of $\Psi_h$ in (\ref{eq:psih}), we have
%
$Q^{(i)} = Q^{(i-1)}\circ \Xi_{h/2}^{-1}\circ \XiL^{-1} \circ \Xi_{h/2}^{-1}$.
%
Map $\Xi_{h/2}(x,v)=(x,v-(h/2)\,\Cov\,\nabla\Phi(x))$ keeps $x$ fixed and translates $v$. Assumption $\nabla_z \Phi(z,\theta)\in \ell_2$
is equivalent to $-(h/2)\,\Cov\,\nabla\Phi(x)$ being an element in the Cameron--Martin space
of the $v$-marginal under $Q_0$, this marginal being $\prod_{i=1}^{\infty}\mathcal{N}(0,1)\otimes
\mathcal{N}(0,A^{-1})$. So, from standard theory for Gaussian laws on general spaces \citep[Proposition 2.20]{prat:92}
 we have
that $Q_0\circ \Xi_{h/2}^{-1}$ and $Q_0$ are absolutely continuous with respect to each
other, with density
\begin{equation}
\label{eq:RN}
G(x,v) =
 \exp\big\{ \langle \tfrac{h}{2}\,g(x), \Cov^{-1/2}v \rangle - \tfrac{1}{2}
  |\tfrac{h}{2}\,g(x)|^2  \big\}\  .
\end{equation}
Assumption $\nabla_z \Phi(z,\theta)\in\ell_2$ guarantees
that all inner products appearing in (\ref{eq:RN}) are finite. Thus,
\begin{align}
\frac{dQ^{(i)}}{dQ_0}(x_i,v_i) &= \frac{d\, \{ Q^{(i-1)}\circ \Xi_{h/2}^{-1}\circ \XiL^{-1} \circ \Xi_{h/2}^{-1}\} }{dQ_0}\,(x_i,v_i)  \nonumber \\
&=
\frac{d\, \{ Q^{(i-1)}\circ \Xi_{h/2}^{-1}\circ \XiL^{-1} \circ \Xi_{h/2}^{-1}\} }
{d\,\{ Q_0\circ \Xi_{h/2}^{-1}  \}}\,(x_i,v_i)\times\frac{d\,\{ Q_0\circ \Xi^{-1}_{h/2}\}}{dQ_0}(x_i,v_i)
 \nonumber
\\
 &=
\frac{d\, \{ Q^{(i-1)}\circ \Xi_{h/2}^{-1}\circ \XiL^{-1}\} }
{dQ_0}\,(\Xi_{h/2}^{-1}(x_i,v_i))\times G(x_i, v_i)\ , \label{eq:tat}
\end{align}
We have  $Q_0\circ \XiL^{-1}\equiv Q_0$,  as
$\XiL$ rotates the infinite-dimensional products
of independent standard Gaussians for the $z,v_z$-components of $Q_0$ and translates the Lebesque measure for the $\theta$-component,
thus overall $\XiL$ preserves $Q_0$.  We also have
$(\XiL^{-1}\circ \Xi_{h/2}^{-1})(x_i,v_i)\equiv \Xi_{h/2}(x_{i-1},v_{i-1})$, so

\begin{align*}
\frac{d\, \{ Q^{(i-1)}\circ \Xi_{h/2}^{-1}\circ \XiL^{-1}\} }
{dQ_0}\,(\Xi_{h/2}^{-1}(x_i,v_i))
&= \frac{d\, \{ Q^{(i-1)}\circ \Xi_{h/2}^{-1}\circ \XiL^{-1}\} }
{d\, \{ Q_0\circ \XiL^{-1}  \}  }\,(\Xi_{h/2}^{-1}(x_i,v_i))     \\
=
\frac{d\, \{ Q^{(i-1)}\circ \Xi_{h/2}^{-1} \} }
{dQ_0}\,(\Xi_{h/2}(x_{i-1},v_{i-1})) 
&= \frac{d Q^{(i-1)} }
{dQ_0}\,(x_{i-1},v_{i-1}) \times
G(\Xi_{h/2}(x_{i-1},v_{i-1}))\ ,
\end{align*}
where for the last equation we divided and multiplied with
$Q_0\circ \Xi_{h/2}^{-1}$, as in the calculations in (\ref{eq:tat}), and used again
(\ref{eq:RN}). Thus, recalling the explicit
expression for $\Xi_{h/2}$, overall we have  that
%
%
\begin{equation*}
\frac{dQ^{(i)}}{dQ_0}(x_i,v_i) = \frac{dQ^{(i-1)}}{dQ_0}(x_{i-1},v_{i-1}) \times
G(x_{i},v_{i})\times G(x_{i-1},v_{i-1}+\tfrac{h}{2}\,\Cov^{1/2}g(x_{i-1}))\ \ .
\end{equation*}
From here one can follow precisely the steps in $\S$ 3.4 of \cite{besk:14} to obtain, for $L=\Cov^{-1}$,
\begin{align*}
\log\{ &\,G(x_{i},v_{i})\, G(x_{i-1},v_{i-1}+\tfrac{h}{2}\,\Cov^{1/2}g(x_{i-1}))\,\} = \\
& =
\tfrac{1}{2}\,\langle x_{i}, Lx_{i} \rangle + \tfrac{1}{2}\,\langle v_{i}, L v_{i}\rangle
- \tfrac{1}{2}\,\langle x_{i-1}, Lx_{i-1} \rangle - \tfrac{1}{2}\,\langle v_{i-1}, L v_{i-1}\rangle\ .
\end{align*}
Thus, due to the cancellations upon summing up, we have proven
the expression for  $(dQ^{(I)}/dQ_0)(x_I,v_I)$ given in statement (i)
of Proposition  \ref{pr:1}.
Given (i), the proof of (ii) follows precisely as in the proof of Theorem
3.1 in \cite{besk:14}.

\bibliographystyle{biometrika}

\bibliography{references}

\begin{thebibliography}{41}
\expandafter\ifx\csname natexlab\endcsname\relax\def\natexlab#1{#1}\fi

\bibitem[{A{\"{\i}}t-Sahalia \& Kimmel(2007)}]{sah:kim2009}
\textsc{A{\"{\i}}t-Sahalia, Y.} \& \textsc{Kimmel, R.} (2007).
\newblock Maximum likelihood estimation of stochastic volatility models.
\newblock \textit{Journal of Financial Economics} \textbf{83}, 413 -- 452.

\bibitem[{Andrieu et~al.(2010)Andrieu, Doucet \& Holenstein}]{Andrieu2010}
\textsc{Andrieu, C.}, \textsc{Doucet, A.} \& \textsc{Holenstein, R.} (2010).
\newblock Particle markov chain monte carlo methods.
\newblock \textit{Journal of the Royal Statistical Society: Series B
  (Statistical Methodology)} \textbf{72}, 269--342.

\bibitem[{Beskos et~al.(2013{\natexlab{a}})Beskos, Kalogeropoulos \&
  Pazos}]{besk:14}
\textsc{Beskos, A.}, \textsc{Kalogeropoulos, K.} \& \textsc{Pazos, E.}
  (2013{\natexlab{a}}).
\newblock Advanced {MCMC} methods for sampling on diffusion pathspace.
\newblock \textit{Stochastic Process. Appl.} \textbf{123}, 1415--1453.

\bibitem[{Beskos et~al.(2013{\natexlab{b}})Beskos, Pillai, Roberts, Sanz-Serna
  \& Stuart}]{besk:13}
\textsc{Beskos, A.}, \textsc{Pillai, N.}, \textsc{Roberts, G.},
  \textsc{Sanz-Serna, J.-M.} \& \textsc{Stuart, A.} (2013{\natexlab{b}}).
\newblock Optimal tuning of the hybrid {M}onte {C}arlo algorithm.
\newblock \textit{Bernoulli} \textbf{19}, 1501--1534.

\bibitem[{Beskos et~al.(2011)Beskos, Pinski, Sanz-Serna \& Stuart}]{besk:11}
\textsc{Beskos, A.}, \textsc{Pinski, F.~J.}, \textsc{Sanz-Serna, J.~M.} \&
  \textsc{Stuart, A.~M.} (2011).
\newblock Hybrid {M}onte {C}arlo on {H}ilbert spaces.
\newblock \textit{Stochastic Process. Appl.} \textbf{121}, 2201--2230.

\bibitem[{Biagini et~al.(2008)Biagini, Hu, {\O}ksendal \& Zhang}]{biag:08}
\textsc{Biagini, F.}, \textsc{Hu, Y.}, \textsc{{\O}ksendal, B.} \&
  \textsc{Zhang, T.} (2008).
\newblock \textit{Stochastic calculus for fractional {B}rownian motion and
  applications}.
\newblock Probability and its Applications (New York). London: Springer-Verlag
  London Ltd.

\bibitem[{Breidt et~al.(1998)Breidt, Crato \& de~Lima}]{BreEtAl1998}
\textsc{Breidt, F.}, \textsc{Crato, N.} \& \textsc{de~Lima, P.} (1998).
\newblock The detection and estimation of long memory in stochastic volatility.
\newblock \textit{Journal of Econometrics} \textbf{83}, 325 -- 348.

\bibitem[{Chib et~al.(2006)Chib, Pitt \& Shephard}]{Chib2006}
\textsc{Chib, S.}, \textsc{Pitt, M.} \& \textsc{Shephard, N.} (2006).
\newblock Likelihood based inference for diffusion driven state space models.
\newblock Working paper.

\bibitem[{Chronopoulou \& Viens(2012{\natexlab{a}})}]{ChrVie2012a}
\textsc{Chronopoulou, A.} \& \textsc{Viens, F.} (2012{\natexlab{a}}).
\newblock Estimation and pricing under long-memory stochastic volatility.
\newblock \textit{Annals of Finance} \textbf{8}, 379--403.

\bibitem[{Chronopoulou \& Viens(2012{\natexlab{b}})}]{ChrVie2012b}
\textsc{Chronopoulou, A.} \& \textsc{Viens, F.} (2012{\natexlab{b}}).
\newblock Stochastic volatility and option pricing with long-memory in discrete
  and continuous time.
\newblock \textit{Quantitative Finance} \textbf{12}, 635--649.

\bibitem[{Comte et~al.(2012)Comte, Coutin \& Renault}]{ComEtAl2012}
\textsc{Comte, F.}, \textsc{Coutin, L.} \& \textsc{Renault, E.} (2012).
\newblock Affine fractional stochastic volatility models.
\newblock \textit{Annals of Finance} \textbf{8}, 337--378.

\bibitem[{Comte \& Renault(1998)}]{ComRen1998}
\textsc{Comte, F.} \& \textsc{Renault, E.} (1998).
\newblock Long memory in continuous-time stochastic volatility models.
\newblock \textit{Mathematical Finance} \textbf{8}, 291--323.

\bibitem[{Cotter et~al.(2013)Cotter, Roberts, Stuart \& White}]{cott:13}
\textsc{Cotter, S.~L.}, \textsc{Roberts, G.~O.}, \textsc{Stuart, A.~M.} \&
  \textsc{White, D.} (2013).
\newblock M{CMC} methods for functions: modifying old algorithms to make them
  faster.
\newblock \textit{Statist. Sci.} \textbf{28}, 424--446.

\bibitem[{Craigmile(2003)}]{crai:03}
\textsc{Craigmile, P.~F.} (2003).
\newblock Simulating a class of stationary {G}aussian processes using the
  {D}avies-{H}arte algorithm, with application to long memory processes.
\newblock \textit{J. Time Ser. Anal.} \textbf{24}, 505--511.

\bibitem[{Da~Prato \& Zabczyk(1992)}]{prat:92}
\textsc{Da~Prato, G.} \& \textsc{Zabczyk, J.} (1992).
\newblock \textit{Stochastic equations in infinite dimensions}, vol.~44 of
  \textit{Encyclopedia of Mathematics and its Applications}.
\newblock Cambridge: Cambridge University Press.

\bibitem[{Deya et~al.(2012)Deya, Neuenkirch \& Tindel}]{deya:12}
\textsc{Deya, A.}, \textsc{Neuenkirch, A.} \& \textsc{Tindel, S.} (2012).
\newblock A {M}ilstein-type scheme without {L}\'evy area terms for {SDE}s
  driven by fractional {B}rownian motion.
\newblock \textit{Ann. Inst. Henri Poincar\'e Probab. Stat.} \textbf{48},
  518--550.

\bibitem[{Dieker(2004)}]{diek:04}
\textsc{Dieker, T.} (2004).
\newblock Simulation of fractional brownian motion.
\newblock MSc Thesis.

\bibitem[{Ding et~al.(1993)Ding, Granger \& Engle}]{DingEtAl1993}
\textsc{Ding, Z.}, \textsc{Granger, C.} \& \textsc{Engle, R.} (1993).
\newblock A long memory property of stock market returns and a new model.
\newblock \textit{Journal of Empirical Finance} \textbf{1}, 83 -- 106.

\bibitem[{Duane et~al.(1987)Duane, Kennedy, Pendleton \& Roweth}]{duna:87}
\textsc{Duane, S.}, \textsc{Kennedy, A.}, \textsc{Pendleton, B.} \&
  \textsc{Roweth, D.} (1987).
\newblock Hybrid {M}onte {C}arlo.
\newblock \textit{Phys. Lett. B} \textbf{195}, 216--222.

\bibitem[{Gatheral et~al.(2014)Gatheral, Jaisson \& Rosenbaum}]{Gath:2014}
\textsc{Gatheral, J.}, \textsc{Jaisson, T.} \& \textsc{Rosenbaum, M.} (2014).
\newblock Volatility is rough.
\newblock Working paper.

\bibitem[{Geyer(1992)}]{geye:92}
\textsc{Geyer, C.} (1992).
\newblock Practical markov chain monte carlo.
\newblock \textit{Statist. Sci.} \textbf{7}, 473--483.

\bibitem[{Girolami \& Calderhead(2011)}]{giro:11}
\textsc{Girolami, M.} \& \textsc{Calderhead, B.} (2011).
\newblock Riemann manifold {L}angevin and {H}amiltonian {M}onte {C}arlo
  methods.
\newblock \textit{J. R. Stat. Soc. Ser. B Stat. Methodol.} \textbf{73},
  123--214.
\newblock With discussion and a reply by the authors.

\bibitem[{Gloter \& Hoffmann(2004)}]{GloHof2004}
\textsc{Gloter, A.} \& \textsc{Hoffmann, M.} (2004).
\newblock Stochastic volatility and fractional brownian motion.
\newblock \textit{Stochastic Processes and their Applications} \textbf{113},
  143 -- 172.

\bibitem[{Golightly \& Wilkinson(2008)}]{goli:08}
\textsc{Golightly, A.} \& \textsc{Wilkinson, D.~J.} (2008).
\newblock Bayesian inference for nonlinear multivariate diffusion models
  observed with error.
\newblock \textit{Comput. Statist. Data Anal.} \textbf{52}, 1674--1693.

\bibitem[{Hosking(1984)}]{hosk:81}
\textsc{Hosking, J. R.~M.} (1984).
\newblock Fractional differencing.
\newblock \textit{Water Resources Research} \textbf{20}, 1898--1908.

\bibitem[{Hu et~al.(2013)Hu, Liu \& Nualart}]{hu:13}
\textsc{Hu, Y.}, \textsc{Liu, Y.} \& \textsc{Nualart, D.} (2013).
\newblock Modified euler approximation scheme for stochastic differential
  equations driven by fractional brownian motions.
\newblock \textit{arXiv preprint arXiv:1306.1458} .

\bibitem[{Jones(2003)}]{Jones2003}
\textsc{Jones, C.} (2003).
\newblock The dynamics of stochastic volatility: evidence from underlying and
  options markets.
\newblock \textit{J. Econometrics} \textbf{116}, 181--224.
\newblock Frontiers of financial econometrics and financial engineering.

\bibitem[{Kalogeropoulos et~al.(2010)Kalogeropoulos, Roberts \&
  Dellaportas}]{kal:rob:del10}
\textsc{Kalogeropoulos, K.}, \textsc{Roberts, G.} \& \textsc{Dellaportas, P.}
  (2010).
\newblock Inference for stochastic volatility models using time change
  transformations.
\newblock \textit{Annals of Statistics} \textbf{38}, 784--807.

\bibitem[{Kou(2008)}]{kou:08}
\textsc{Kou, S.~C.} (2008).
\newblock Stochastic modeling in nanoscale biophysics: subdiffusion within
  proteins.
\newblock \textit{Ann. Appl. Stat.} \textbf{2}, 501--535.

\bibitem[{Kou \& Xie(2004)}]{kou:04}
\textsc{Kou, S.~C.} \& \textsc{Xie, X.~S.} (2004).
\newblock Generalized langevin equation with fractional gaussian noise:
  Subdiffusion within a single protein molecule.
\newblock \textit{Physical Review Letters} \textbf{93}, 180603.

\bibitem[{Lobato \& Savin(1998)}]{LobSav1998}
\textsc{Lobato, I.~N.} \& \textsc{Savin, N.~E.} (1998).
\newblock Real and spurious long memory properties of stock market data.
\newblock \textit{Journal of Business and Economic Statistics} \textbf{16},
  261--268.

\bibitem[{Lysy \& Pillai(2013)}]{lysy:13}
\textsc{Lysy, M.} \& \textsc{Pillai, N.} (2013).
\newblock Statistical inference for stochastic differential equations with
  memory.
\newblock Tech. rep.

\bibitem[{Mandelbrot \& Van~Ness(1968)}]{mand:68}
\textsc{Mandelbrot, B.~B.} \& \textsc{Van~Ness, J.~W.} (1968).
\newblock Fractional {B}rownian motions, fractional noises and applications.
\newblock \textit{SIAM Rev.} \textbf{10}, 422--437.

\bibitem[{Mishura(2008)}]{mish:08}
\textsc{Mishura, Y.~S.} (2008).
\newblock \textit{Stochastic calculus for fractional {B}rownian motion and
  related processes}, vol. 1929 of \textit{Lecture Notes in Mathematics}.
\newblock Berlin: Springer-Verlag.

\bibitem[{Norros et~al.(1999)Norros, Valkeila \& Virtamo}]{norr:99}
\textsc{Norros, I.}, \textsc{Valkeila, E.} \& \textsc{Virtamo, J.} (1999).
\newblock An elementary approach to a {G}irsanov formula and other analytical
  results on fractional {B}rownian motions.
\newblock \textit{Bernoulli} \textbf{5}, 571--587.

\bibitem[{Prakasa~Rao(2010)}]{rao:10}
\textsc{Prakasa~Rao, B. L.~S.} (2010).
\newblock \textit{Statistical inference for fractional diffusion processes}.
\newblock Wiley Series in Probability and Statistics. Chichester: John Wiley \&
  Sons Ltd.

\bibitem[{Roberts \& Stramer(2001)}]{rob:str01}
\textsc{Roberts, G.} \& \textsc{Stramer, O.} (2001).
\newblock On inference for partial observed nonlinear diffusion models using
  the metropolis-hastings algorithm.
\newblock \textit{Biometrika} \textbf{88}, 603--621.

\bibitem[{Rosenbaum(2008)}]{Ros2008}
\textsc{Rosenbaum, M.} (2008).
\newblock Estimation of the volatility persistence in a discretely observed
  diffusion model.
\newblock \textit{Stochastic Processes and their Applications} \textbf{118},
  1434 -- 1462.

\bibitem[{Stramer \& Bognar(2011)}]{str:bog11}
\textsc{Stramer, O.} \& \textsc{Bognar, M.} (2011).
\newblock Bayesian inference for irreducible diffusion processes using the
  pseudo-marginal approach.
\newblock \textit{Bayesian Analysis} \textbf{6}, 231--258.

\bibitem[{Sussmann(1978)}]{suss:78}
\textsc{Sussmann, H.~J.} (1978).
\newblock On the gap between deterministic and stochastic ordinary differential
  equations.
\newblock \textit{Ann. Probability} \textbf{6}, 19--41.

\bibitem[{Wood \& Chan(1994)}]{wood:94}
\textsc{Wood, A.} \& \textsc{Chan, G.} (1994).
\newblock Simulation of stationary {G}aussian processes in $[0,1]^d$.
\newblock \textit{J. Comput. Graph. Statist.} \textbf{3}, 409--432.

\end{thebibliography}

\end{document}